\documentclass[preprint]{Jennifers-aastex}  
\usepackage{geometry}                		
\usepackage[parfill]{parskip}    		
\usepackage{graphicx}				
\usepackage{amssymb}
\usepackage{appendix}

\title{The National Science Foundation's AST Portfolio Review of 2012  is Not Relevant 
to the Green Bank Telescope of 2017: A White Paper
}

\author{Felix J. Lockman, Ryan Lynch, David T. Frayer}
\affil{Green Bank Observatory\altaffilmark{1}, Green Bank, WV 24944}

\author{Brian D. Mason,  Scott M. Ransom}
\affil{National Radio Astronomy Observatory\altaffilmark{1}, Charlottesville, VA 22901}

\altaffiltext{1}{The Green Bank Observatory, and the National Radio Astronomy Observatory, are operated by Associated Universities, Inc., under a cooperative agreement with  the National Science Foundation.}

\begin{document}
\maketitle

\section*{Summary}
 The National Science Foundation (NSF) Astronomy Division's Portfolio Review of 2012 \cite{Eisenstein}  
 is no longer relevant to the 
 Green Bank Telescope (GBT) 
 of 2017 for two principal reasons, one instrumental and the other astrophysical: 1) The GBT has 
 begun significant operations in the 3mm band,  giving it unrivaled capabilities for spectroscopy and continuum studies over 67-116 GHz.  It is now an instrument that is unique worldwide and 
is a critical complement to ALMA for the U.S. scientific community.  These capabilities had not 
been implemented at the time of the review.
2) The detection of gravitational radiation by LIGO in 2015 
places the GBT's work on pulsar observations of nano-Hz gravitational radiation at the forefront of modern astrophysics.  

The Green Bank Telescope of 2017  is not the GBT that was reviewed by the Eisenstein-Miller committee in 2012, 
a review that was specific to the NSF Astronomy portfolio \cite{Eisenstein}.  
The GBT serves a wide spectrum of science areas including physics, chemistry, and planetary studies as well 
as astronomy.  Besides  its well-documented intellectual merit, it thus has a significant broader impact.  
The GBT is making significance advances in our understanding of 
gravitational waves, the equation-of-state of nuclear matter, the mass of  supermassive black holes, 
the value of  ${\rm H_0}$, and  the physics of star-formation,  all key science goals for astronomy identified in a recent 
National Academy study {\it New Worlds, New Horizons: A Midterm Assessment}
 \cite{NatAcad}.    In the era of ALMA and LIGO,  
  other countries have bolstered their mm-wave and cm-wave facilities;
it is critical that U.S. scientists have ready access to a large filled aperture to remain at the forefront of 
research. 

\section{BEFORE THE PORTFOLIO REVIEW: 2001 -- 2011}

\subsection{GBT Instrumental Capabilities}

The Green Bank Telescope (GBT) is one of the newest facilities of the Division of Astronomical Sciences (AST) of the 
National Science Foundation (NSF).	At the beginning of its scientific operations in 2001, the GBT was unique in having 
an active surface that could be used to remove gravitational and thermal distortions  in real time 
\cite{Prestage}.  A surface correction of this type had never before been attempted on a radio telescope 
\cite{Hughes}, so its implementation was not part of the construction contract but was 
left for the NRAO staff to achieve sometime within the first decade of GBT operations 
\cite{Hall}.   With a passive surface the GBT has acceptable performance at frequencies up to 15 GHz, 
but it was understood that use of the active surface would allow operations in the 3mm atmospheric window 
up to 116 GHz \cite{Hughes}, where the Green Bank site has good atmospheric conditions for more than 2000 hours a year \cite{Lockman2010}.    
Because the style in the United States is to fund telescope construction fully while underfunding the transition to operations 
\cite{Zimmerman}, it was not until 2010 that the GBT surface panels were adjusted precisely and the 
active surface was completely operational \cite{Hunter}.   At that time the telescope pointing and tracking were also improved.

	The GBT construction budget included funding for  
a suite of receivers covering 0.2 - 50 GHz based on designs from the 
1990's \cite{Balister,Behrens}, and a spectrometer whose design was established at the same time 
\cite{Lockman1992,Langston,Escoffier}.   
In 2010, when it finally became possible to use the telescope at 3mm with good efficiency,  there was no  funding for receivers to operate in this band, so 
by the start of 2012, 
GBT operations in the 3mm band were limited to the pathfinder MUSTANG instrument, 
a 64-pixel bolometer array built by a group at the University of Pennsylvania \cite{Dicker2008}.  
It confirmed the GBT's performance at 90 GHz and produced significant results on  
 the formation of galaxy clusters \cite{Mason}, but had a relatively small field of view that limited its 
general utility,  and  no spectroscopic capabilities.  

It was at this time that the NSF's Eisenstein-Miller  Portfolio Review occurred \cite{Eisenstein}. 

\subsection{ The State of Gravitational Wave Astronomy Back in 2012}

Gravitational Waves (GW) can be created by a number of sources, and the GW spectrum  spans many orders 
of magnitude in frequency,  much like the electromagnetic spectrum.  
 GWs in the nanohertz (nHz) frequency range  are expected to arise from supermassive binary black holes  in the early stages of inspiral, and potentially exotic sources such as cosmic strings. They can be
 detected through variations in the arrival time of pulses from millisecond radio pulsars \cite{Lommen,Sanidas}. 

Because of its large collecting area, broad sky coverage, sensitive receivers, and location in the National Radio 
Quiet Zone,  the GBT is the world's premier telescope for the study of pulsars and other compact objects (see \S4).  
Together with the Arecibo Observatory, the GBT is crucial to the 
North American Nanohertz Observatory for Gravitational Waves (NANOGrav), an   
NSF-funded North American collaboration working towards a direct detection of nHz GWs \cite{Ransom16}.  
This complements ground-based  GW interferometers in the kilohertz window and possible future space-based interferometers in the millihertz window.  Pulsar timing arrays are also sensitive to GW bursts with memory, a key but unproven prediction of general relativity \cite{CordesJenet}.

	In 2012, at the time of the NSF's Eisenstein-Miller Portfolio review, direct detection of the GW universe was not 
possible (see Figure 1).  The upper limits on GW amplitude being set by
pulsar timing measurements were higher than the amplitude predicted by most theoretical models \cite{Cordes}, and 
initial LIGO operations had concluded without a detection.  
Advanced LIGO was only beginning to be installed \cite{LIGO}.

\section{SINCE THE PORTFOLIO REVIEW: 2012 -- 2017}

\subsection{The GBT High-frequency Receiver Development}

	In early 2012 a 7-pixel receiver for spectroscopy in the 
18--26 GHz band (built primarily using non-Federal funds) was put into regular service, 
and in the spring of 2012 a receiver was also installed to work at the lower end of the 3mm window  
from 67--92 GHz.  
This latter ``W-band'' receiver has only two pixels, too widely spaced for efficient 
mapping, and was built from existing parts with a modest upgrade in late 2012. 

	Because of the lack of funding for GBT receiver development
in the NRAO operations budget, it is only now, in 2016, that receivers have 
become available to exploit some of the GBT's new capabilities. 
The new receivers are Argus and MUSTANG-2.

	Argus is a 16-pixel array for spectroscopy at 75--116 GHz built by a consortium led by 
Sarah Church at Stanford \cite{Sieth}.
MUSTANG-2 is a 211 pixel bolometer array with a 30 GHz bandwidth around 90 GHz, built by a consortium led by 
Mark Devlin at the University of  Pennsylvania \cite{Dicker2014}.  
Both were funded through separate competitions in the NSF AST-ATI program. 
Argus is a testbed for new scalable receiver technologies, while MUSTANG-2 builds on the experience with the earlier bolometer array, MUSTANG, with a significantly improved design.  
In support of Argus and other GBT receivers  there is a new spectrometer, VEGAS,
 built by a consortium  led  by Dan Wertheimer at University of California, Berkeley  \cite{Chennamangalam}, 
again funded through a successful competition in the NSF AST-ATI program.

	Argus gives the GBT spectroscopic capabilities in the molecule-rich upper frequencies of the 3mm band.  
Figure 2 shows the first Argus map from the complete 16-pixel array.  This $^{13}$CO map 
was made in only 40 minutes in September 2016 in poor weather (zenith $\tau=0.42$ at 110 GHz).  Nonetheless,
 it demonstrates the unique power of the GBT for high angular resolution wide-field mapping.  
  Argus is heavily oversubscribed and will be in regular use during the 2016--2017 winter season for projects 
described in $\S3.1$.

\begin{figure}
\scalebox{0.40}[0.40]{\includegraphics[angle=-0]{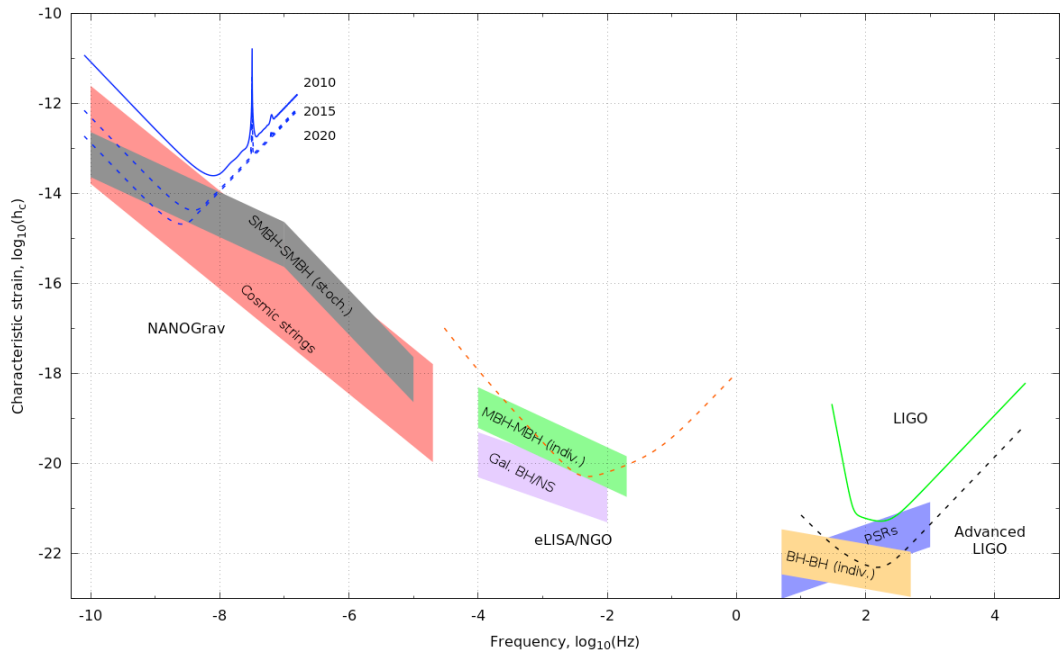}}
  \caption
  {This figure, from 2012, shows the state of gravitational wave astronomy at that time.  The colored areas 
  show theoretical expectations, and the solid and dashed lines the predicted sensitivity of various experiments. 
  Both Advanced LIGO and NANOGrav have achieved their predicted sensitivities, but in the case of LIGO 
  the theoretical expectations were correct, while for NANOGrav they were not \cite{Abbott,Arzoumanian}.  
  Newer models predict that a direct detection of nHz gravitational waves is likely within 5 years \cite{Taylor}.  
  eLISA/NGO is not operational. {\it Figure courtesy of Paul Demorest.}
}  
\label{fig:GW_power}
\end{figure}

\begin{figure}
\begin{center}
\scalebox{0.5}[0.5]{\includegraphics{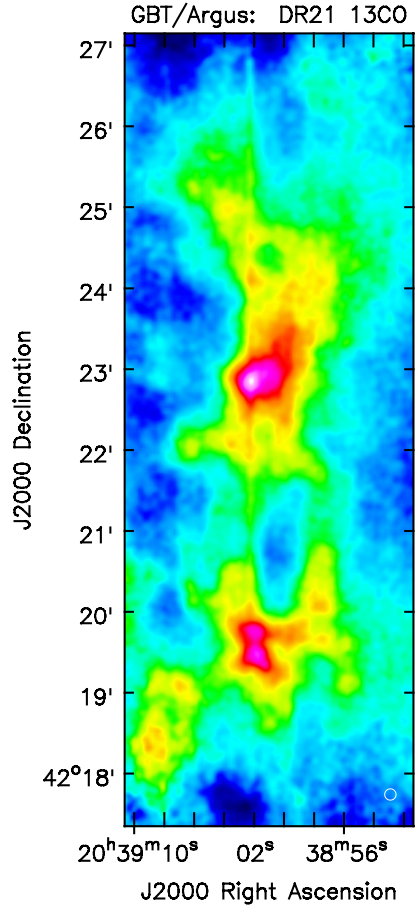}}
  \caption
  {Argus map of $^{13}$CO around DR 21 made during instrument commissioning in 
  September 2016.    
  Emission is integrated over 10 km s$^{-1}$.
  The angular resolution is shown by the small circle at the lower right.   
  The map was made in 40 minutes on the GBT,  in poor weather, 
  with a zenith opacity at 110 GHz of $\tau=0.42$.
}  
\label{fig:Argus}
\end{center}
\end{figure}

	MUSTANG-2 has about 4 times more pixels than the original MUSTANG, each with a greater sensitivity. 
 With its wide field of view and better stability, it should be at least 30 times faster.  
It will be commissioned in the 2016-2017 winter observing season.

	The GBT has not yet reached its full potential in the 3mm band.
Daytime thermal distortions of the surface and pointing degradation from wind currently 
limit 3mm operations to calm nights (Appendix A).   
However, there are straightforward solutions to these limitations that can be achieved at 
relatively modest cost ($\lesssim \$1$M).  If such funding became available  
 the useable observing time at 3mm could be increased by a factor $\sim 2$ \cite{Prestage2015}.

\subsection{The Changing Landscape of Gravitational Wave Astronomy}

The spectacular discovery by LIGO of multiple GW 
sources in the $\sim100$ Hz frequency band  
has officially ushered in the era of GW astronomy \cite{Abbott},  and 
pulsar timing still continues to be the only tool available for studying the GWs from the evolution of 
super massive Binary Black Holes (SMBBHs). 
The NANOGrav collaboration has been achieving its predicted sensitivity and is probing 
interesting GW phase space, placing severe constraints on the simplest models of SMBBH mergers 
\cite{Arzoumanian}.  
The latest NANOGrav results are also beginning to provide independent constraints on parameters 
such as the eccentricity distribution of SMBBHs or the coupling between SMBHHs and their local galactic 
environments \cite{Arzoumanian}.

	While the amplitude of the stochastic nHz GW background now appears less than predicted by the early 
models, unlike the GW 
events at $\sim100$ Hz probed by LIGO which are appearing at  expected levels
(Fig.~1), newer
models  predict that a direct detection of nHz GWs is likely within 5 years 
and almost certain in 10 years, 
if the GBT (and Arecibo) continue at their current levels of sensitivity \cite{Taylor}.  
The GBT is especially important because of its extensive sky coverage 
which is necessary to detect the correlated signals expected from the GW background.  The high 
aperture efficiency of the GBT, its very sensitive receivers, and relative freedom from interfering signals 
 make it 2 to 4 times 
faster for pulsar timing than the JVLA \cite{Ransom16} (see also \S3.3).

\section{THE ROLE OF THE GBT IN US SCIENCE:  2017 ONWARDS }

\subsection{Spectroscopy at 3mm: A Critical Complement to ALMA}

The new Argus camera is demonstrating the unique capabilities of the GBT in the 
 3mm atmospheric window, which contains a very large number of important spectral lines including the 
ground-state transitions of CO and its isotopologues, as well as HCO$^+$,  SiO,  HNC, N$_2$H$^+$, CS and HCN.    
The scientific significance of this band has been documented exhaustively 
(e.g., \cite{Fuller,Meier}).  
The GBT now offers the high sensitivity of a filled aperture with an angular resolution of 
$6\arcsec$-$10\arcsec$,  
well-matched to the mid- and far-infrared observations from the Spitzer, SOFIA, and Herschel telescopes.
The GBT is ideal for resolving cold cores within local star-forming regions, 
tracing the dynamics of molecular filaments, 
and for mapping giant molecular clouds in nearby galaxies.  
Its powerful wide-field mapping capabilities for 3mm spectroscopy are unique, and unrivaled worldwide.  
 The GBT provides a wide-field complement to ALMA and can supply critical short-spacing data \cite{Caldu-Primo} for 3mm synthesis observations.
 
\begin{deluxetable}{ccccccc}
\tablecaption{{Time to map spectra at 86 GHz over a $3^{\prime} \times 3^{\prime}$ field to 20 mK rms\tablenotemark{a}}
\label{tab:spectral_mapping}}
\tablehead{
\colhead{GBT 2015} &\colhead{Argus 2015} & \colhead{GBT 2020}  & \colhead{ALMA} & \colhead{ALMA}  & \colhead{ALMA}  &  \colhead{ngVLA} \\
\colhead{1 Pixel} &  \colhead{16 Pixels} & \colhead{50 Pixels} & \colhead{50x12m}  &  \colhead{50x12m} & 
\colhead{10x7m}  & \colhead{300x18m}   \\
\colhead{$8^{\prime\prime}$} & \colhead{$8^{\prime\prime}$} & \colhead{$8^{\prime\prime}$} & \colhead{$3^{\prime\prime}$} & \colhead{$1^{\prime\prime}$}  & \colhead{$23^{\prime\prime}$}  &  \colhead{$1^{\prime\prime}$}
}
\startdata
21h & 3.3h     &   $<1$h         & 19h  &  1,500h & 0.5h & 17,500h 
\enddata
\tablenotetext{a}{\small For $\delta=0\arcdeg$, 1 km s$^{-1}$ channels, GBT Ta* 2000 h/yr opacity, 
ngVLA tapered to 1\arcsec\  resolution at 80 GHz for the Clark/Conway configuration \cite{Carilli}.  GBT2020 has dual polarization.  ALMA numbers from the ALMA sensitivity calculator on 10 July 2016. No allowances for overhead.}
\end{deluxetable}

	 Table 1 compares the time it would take to reach a Ta* noise limit of 20 mK in a 
$3\arcmin \times 3\arcmin$ field for the current single-pixel receiver on the GBT, for  
the 16-pixel Argus, and for a 50-pixel receiver that could be built with existing technology, in comparison with ALMA  and the proposed ngVLA \cite{Carilli}.  
Green Bank has acceptable observing conditions for 3mm spectroscopy for more than 2000 hours each year \cite{Lockman2010}, during which 
 the GBT's 86 GHz receiver has a system temperature at zenith about 
equal to that of ALMA (Appendix A).  It is clear that the GBT occupies a critical place in U.S. 3mm instrumental capabilities.

The GBT at 3mm is now making contributions to areas as diverse as the distribution of 
dense molecular gas in nearby galaxies and the 
structure of the  jet of  M87 \cite{Kepley,Hada}.  
General science areas where it will contribute in coming years include interstellar chemistry, comets, the context of star formation, stellar outflows, Galactic molecular clouds and complexes, molecules in nearby galaxies, and molecules in galaxy clusters. 

	Approved Argus projects for the GBT winter observing of 2016--2017 include: 
\begin{itemize}
  \item	Measurement of the physical conditions and kinematics in star-forming filaments;
  \item Surveys  of the infall and fragmentation in dense pre-stellar cores;
  \item Study of the transition between atomic and molecular gas in the Galactic halo; 
  \item Tests of theories of  fragmentation in interstellar filaments;   
\item Investigation into the formation paths of organic molecules in protostars; 
\item Characterization of the distribution of dense molecular gas in nearby galaxies.
\end{itemize}

	The technical path to a 50 pixel camera for the GBT at 3mm is straightforward and follows from 
existing technology and precursors like Argus.  
Such an instrument would transform our understanding of many fields, 
open up new areas of research, and, at relatively modest cost, add significant capability to 
the ensemble of U.S. astronomical facilities (Table 1).  
The focal plane of the GBT can support 3mm cameras with many hundreds of pixels without significant coma or gain loss \cite{Lockman2016}. 
Phased-array-feed technology, though still in engineering tests at this time, may lead to multi-pixel receivers for highly efficient spectral line mapping in this band \cite{Erickson}.

\subsection{Ultra-sensitive 3mm Continuum Studies: Dust and the SZE}

The pathfinder MUSTANG showed that the GBT was a unique instrument for 90 GHz 
continuum studies which include high angular-resolution 
mapping of the Sunyaev-Zel'dovich Effect (SZE) in galaxy clusters \cite{Mason, Korngut, Mroczkowski} and 
measuring the physical properties of interstellar dust \cite{Schnee}.  MUSTANG-2 is  
expected to have a  factor of 30  improvement in speed as well as increased stability.  
Table~2 compares the wide-field continuum mapping capabilities of MUSTANG-2 with ALMA and the ngVLA. Note that neither Table~1 nor Table~2 accounts for the time needed to obtain the 
short-spacing data that must be added to 
the interferometer observations to restore all the flux in the field.  

	MUSTANG-2 will produce a high-resolution ($9^{\prime\prime}$) $62\  \mu$K rms image of 
the SZE in a galaxy cluster in just one hour.  
In the course of a winter observing season it could image hundreds of massive clusters.  
In 2 hours it will measure the pressure profiles in the inter-cluster medium to a radius of 
$3^{\prime}$ in systems as small as $4.5\times 10^{14} \  M_{\odot}$.  It can image the cavities found 
in X-ray emission from clusters much faster than ALMA while recovering emission on scales larger than ALMA's $\sim 1\arcmin$ field of view at 3mm.

MUSTANG-2 will be a science pathfinder for ALMA and JWST,  
mapping large areas to identify  sources for 
studies of galaxy formation in the high-redshift Universe.  
Its can detect $\sim 5000$ sources per square degree down to 
the confusion limit, providing constraints on star formation rates at $z>5$.
For Galactic studies, in just a few hundred hours on the GBT
MUSTANG-2 can make a 10,000 deg$^2$ survey of the Galaxy to a 
$5\sigma$ completeness limit of 8 mJy, 
complementing surveys from Herschel, Bolocam, and Planck.  
As Table~2 shows, the GBT is a powerful and unique  instrument for continuum studies at 3mm.

\begin{deluxetable}{ccccc}
\tablewidth{5.7in}
\tablecaption{{3mm Continuum rms noise over a  $5\arcmin \times 5\arcmin$ field in 1 hour\tablenotemark{a}}
\label{tab:continuum_mapping}}
\tablehead{
\colhead{GBT} & \colhead{ALMA}  & \colhead{ALMA} & \colhead{ALMA}  & 
\colhead{ngVLA} \\
\colhead{MUSTANG-2}  & \colhead{50x12m} & \colhead{50x12m} &  \colhead{10x7m}  & \colhead{300x18m}    \\
\colhead{$9^{\prime\prime}$} & \colhead{$3^{\prime\prime}$} & \colhead{$1^{\prime\prime}$}  & 
\colhead{$23^{\prime\prime}$}  &  \colhead{$1^{\prime\prime}$}  
}
\startdata
 $62\  \mu$Jy  & $49\  \mu$Jy   & $49\  \mu$Jy  & $1.35\ $mJy & $65\  \mu$Jy  \\
 0.062 mK\tablenotemark{b} & 0.9 mK & 8.0 mK & 0.12 mK &  13.0 mK 
\enddata
\tablenotetext{a} {\small ngVLA values for a 30 GHz bandwidth at 80 GHz tapered to $1^{\prime\prime}$ resolution \cite{Carilli}; 
ALMA for 7.5 GHz band around 86 GHz; 
MUSTANG-2 for 75-105 GHz \cite{Dicker2014}.   No allowances for overhead.}
\tablenotetext{b} {\small This is ${\rm T_a^*}$.  To convert to ${\rm T_{mb}}$ use a factor 1.25 (including stable error beam) or 1.74 (main beam only).}
\end{deluxetable}

\subsection{Gravitational Radiation and Fundamental Physics: 2017 Onwards}

The GBT will continue to be the premier telescope for pulsar astronomy because of its sensitivity, sky coverage, 
and relative freedom from RFI afforded by the unique National Radio Quiet Zone.  
In a direct comparison, the GBT proved to be from 2 to 4 times faster than the JVLA for pulsar timing
despite the larger collecting area of the JVLA \cite{Ransom16}.  
Searching for new pulsars using the JVLA is not practical because of extreme computational complexity.

Ongoing surveys and follow-up of unidentified Fermi sources 
are discovering $\sim4$ new millisecond pulsars a year that are being added to the NANOGrav sample \cite{Ransom16}.
As the sensitivity to GWs is proportional to the number of pulsars in the timing array, 
the steady rate of discovery is one of the main reasons that NANOGrav is achieving 
its predicted improvements in sensitivity.  
A new ultra-wideband receiver ($\sim0.7 - 4$ GHz)   
could improve NANOGrav sensitivity on the GBT by as much as a factor of two 
by reducing the uncertainties arising from interstellar scattering. 
This receiver would also be extremely useful for studies of transient sources such as 
Fast Radio Bursts  and other pulsar timing experiments that probe extreme physics, an area 
where the GBT is beginning to contribute \cite{Masui}. 

	GW astronomy was  unproven in 2012, but in 2017 is a fully realized 
observational window with demonstrable results and paths for improvements in sensitivity.  
The GBT is at the forefront of the direct detection of nHz GWs in the next decade, but detection is only the first step.  
Pulsar timing arrays like NANOGrav are informing our knowledge of SMBBH mergers 
including the solution to the ``final parsec'' problem \cite{Milosavljevic,Sampson}, and will continue to do so.  
As the sensitivity of NANOGrav improves, 
it could measure the properties of individual 
SMBBH systems and anisotropies in the GW background \cite{Mingarelli}.
These measurements are also the best way of testing the existence of cosmic strings \cite{Sanidas}; 
the latest constraints significantly surpass the limits from the Planck satellite \cite{Arzoumanian}.

\section{COMPLEMENTARY CURRENT GBT SCIENCE}

In addition to the science described above, 
 the GBT continues to be significantly oversubscribed \cite{TAC_16A,TAC_16B}
   for a broad range of research that takes advantage of its unique capabilities at frequencies below 50 GHz.  
     The following list 
 gives a representative sample of recent GBT research that cuts across many disciplines:

	Chemistry ---  With its high sensitivity to low surface-brightness, the GBT has been used to discover 
$\approx20$  new interstellar organic molecules including C$_6$H$^-$, the first 
interstellar anion \cite{McCarthy}, and Propylene Oxide, the first interstellar chiral molecule \cite{McGuire}.
	
	Solar System ---  The GBT is used regularly as the receiving element of bi-static radar programs to study  Lunar geology \cite{Campbell}, the rotation state of planets and their satellites 
\cite{Margot2013}, and the structure and orbit  of near-Earth objects  \cite{Asteroid}.   
One program reported conclusive evidence for a liquid core in the planet Mercury \cite{Margot2007}
	
	Star formation --- GBT observations often serve as the pathfinder  for 
higher-frequency or higher-resolution observations with the JVLA  and ALMA.
	The GBT has studied the evolution of dense cores in 
filamentary clouds \cite{Seo}, and  determined their evolutionary state through 
abundance changes in long carbon-chain molecules \cite{Friesen}.  Wide-area mapping 
 in transitions of NH$_3$ has 
 measured the gas dynamics, temperature, and chemistry in star-forming 
regions \cite{Sadavoy}, and 
identified areas where turbulence dissipates prior to star formation  \cite{Pineda}.  
More than 1,000 new Galactic HII regions have been discovered \cite{Anderson}.

Stars --- The GBT provided the critical sensitivity for VLBI parallax measurements that 
     determined  an accurate distance to the Pleaides \cite{Melis}.  A companion star to a pulsar studied with the 
 GBT is the coolest  known white dwarf, with an implied age $\sim 10$ Gyr \cite{Kaplan}.
	
	Galaxies --- The GBT has made extremely sensitive observations of HI in dwarf galaxies \cite{Spekkens},
measured very faint neutral circumgalactic gas \cite{Wolfe}, and  
detected H$_2$O masers in the disk of M31 \cite{Darling}.  GBT surveys of HI in galaxies 
revealed the structure of the local Universe 
\cite{Tully}.  The GBT has discovered H$_2$O megamasers around galactic nuclear accretion disks 
\cite{Gao}, probed the physics of accretion disks \cite{Pesce}, and measured the mass of numerous 
supermassive black  holes with errors below one percent \cite{Gao, Greene}.
Megamasers give an 
independent measurement of the Hubble constant from simple geometry \cite{Braatz}.
	
	Fundamental physics --- The GBT is used to place limits on violations of the equivalence principle  
\cite{Ransom},  to limit  variations in the gravitational 
constant G outside the Solar System \cite{Zhu},  and to  measure general relativistic effects in the 
strong field approximation \cite{Breton}.  Discovery of a 2M$_{\sun}$ pulsar with the GBT 
limits the equation of state of matter in its densest form \cite{Demorest,Ozel}. 
 The importance
of the GBT to the nuclear astrophysics community was emphasized in a recent white paper \cite{Whitepaper}.

\section{THE GBT AND THE U.S. SCIENTIFIC COMMUNITY}

The GBT serves a large and diverse scientific community who work  in astronomy, 
chemistry, planetary science, physics, STEM education, and related fields \cite{Comets}.   

The GBT provides critical capabilities not otherwise available to U.S. scientists, and 
 is in high demand \cite{TAC_16A,TAC_16B}.
In the three years 2013--2015 the telescope was used by 
more than 950 individual scientists and their students.  
A typical GBT proposal in that period had five investigators, one of whom was a student, 
and one of whom was from an institution outside the U.S.
The user community is growing.  
Nearly 750 individuals proposed  to use the GBT in 2016, 337 of whom had not 
used the GBT in 2013--2015.  This is a substantial increase from the 600 individuals who 
proposed to use the GBT in 2012.   
Its new capabilities at 3mm ensure a continued growth in the GBT user community. 

The GBT is a unique facility available to scientists for developing  new instrumentation 
\cite{Chennamangalam,Dicker2008, Dicker2014, Erickson, Harris, Sieth} 
and new astronomical techniques (e.g.,  
 HI intensity mapping \cite{Chang}).  It also offers 
 opportunities for hands-on training in astronomical observing methods
 [e.g., NRAO eNews 5, 7], an increasingly 
 rare resource at all wavelengths  \cite{OIR}.  While other nations are bolstering their mm-wave and cm-wave 
 facilities in response to ALMA and LIGO \cite{NOEMA,FAST}, the U.S. is closing or reducing access to its facilities.

The GBT is integrated into a broad program of Education and Public Outreach 
to the more than $40,000$ visitors who come to the Green Bank Observatory each year.  Programs include science teacher training, hands-on use of radio telescopes by  K-12 students and their teachers, programs for amateur astronomers, 
visiting college professors and tourists, among others.  
The Pulsar Search Collaboratory has involved more than 2,000 high school students and  
more than 100 teachers in the analysis of GBT data resulting in the discovery of seven new pulsars \cite{PSC,Rosen}.
The GBT is a rare public showcase for NSF-funded science.

\section{ CONCLUSION}

The National Academies study {\it New Worlds, New Horizons: A Midterm Assessment}
 \cite{NatAcad}  highlighted the importance of measuring gravitational waves, 
 constraining the equation-of-state of nuclear matter, determining the mass of  
 supermassive black holes, determining  ${\rm H_0}$, and understanding the physics 
 of star-formation,  as key science goals for astronomy.   All of these key areas are currently being advanced using the GBT,  
e.g., \cite{Arzoumanian,Demorest, Greene, Gao, Braatz, Pineda, Friesen,Seo}.  

	The NSF's Astronomical Sciences Portfolio Review of 2012 has been overtaken by subsequent events, 
both scientific and instrumental.   Its conclusions with respect to the Green Bank Telescope are no longer 
relevant to the needs of the U.S. scientific community in 2017.

\section{Acknowledgements}
We thank many colleagues who provided input and advice, including   Loren Anderson, Dana Balser, Tom Bania, Jim Braatz,  Joel Bregman, Nichol Cunningham, Jeremy Darling, Frank Ghigo, Paul Goldsmith, Amanda Kepley,  Natalia Lewandowska, Dunc Lorimer, Ron Maddalena,  Maura McLaughlin, Toney Minter,  D.J. Pisano and Richard Prestage.



\section*{Appendix A:  Atmospheric Conditions at Green Bank for 3mm Observations}

Atmospheric opacity is the fundamental limit on use of the GBT in the 3mm band, and results from 
many years of monitoring are presented in	GBT memo 267 \cite{Lockman2010}.
 In summary, at 86 GHz there are extensive periods of acceptable atmospheric opacity 
($\tau<0.2$) throughout the entire year, but especially in the winter observing season between 
1 October and 1 May.  During this time there are typically 1300 hours for which 
$\tau<0.1$, and 2000 hours for which $\tau<0.14$.  
These periods are predictable several days in advance.  
GBT data in Table 1 are for $\tau=0.14$.  
(GBT data in Table 2 are derived from experience at the GBT with MUSTANG.) 
The opacity rises rapidly toward the lower and higher frequency edges of the 3mm band because of absorption by Oxygen.  
	
	There are also considerations of wind.  Most nights during the winter season have a mean wind speed low enough to permit accurate GBT measurements 
throughout the 3mm band with the current GBT optics and control systems.   
All in all, there are currently $\sim1000$ hours of ``photometric'' quality conditions 
($\tau<0.1$ at 86 GHz) during each winter season that also have low wind-induced pointing errors.   
Wind-induced errors could be reduced by improvements to the GBT optics and servo systems.

	At 86 GHz the current single-pixel GBT receiver has a receiver temperature $\sim50$ K, 
and an atmosphere with a zenith opacity $\tau=0.1$ adds 27 K of noise, so counting all contributions, 
the zenith Tsys is $\sim85$ K for $>1000$ hours each year, a Tsys comparable to 
that of ALMA at this frequency.

\end{document}